\documentclass[12pt]{article}
\pdfoutput=1
\usepackage{amsmath,amsfonts,graphicx,color,bbm,tikz}
\usepackage{comment}
\usepackage[nosort]{cite}
\usepackage{subfigure}
\usetikzlibrary{calc,positioning}
\usetikzlibrary{patterns,arrows,decorations.pathreplacing}
\usepackage[width=.75\textwidth]{caption}
\usepackage{ulem}
\tikzset{>=stealth}

\makeatletter\@addtoreset{equation}{section}\makeatother

\newcommand{\be}{\begin{equation}}
\newcommand{\ee}{\end{equation}}
\newcommand{\bea}{\begin{eqnarray}}
\newcommand{\eea}{\end{eqnarray}}

\newcommand{\bra}[1]{{\left< {#1} \right|}}
\newcommand{\ket}[1]{{\left| {#1} \right>}}

\newcommand{\binomi}[2]{\begin{pmatrix} #1 \\ #2 \end{pmatrix}}

\renewcommand{\title}[1]{\vbox{\center\LARGE{#1}}\vspace{3mm}}
\renewcommand{\author}[1]{\vbox{\center#1}\vspace{3mm}}

\newcommand{\email}[1]{\vbox{\center\tt#1}\vspace{3mm}}

\begin{document}

\rightline{\small{\tt }}
\begin{center}

\vskip-1.5cm
{\large {\bf Non-Interacting Motzkin Chain - Periodic Boundary Conditions} }
\vskip 0.75cm

Olof Salberger$^\dagger$, Pramod Padmanabhan$^*$ and Vladimir Korepin$^\dagger$, 

\vskip 0.5cm 

${}^\dagger $ C.N.Yang Institute for Theoretical Physics, State University of New York, Stony Brook, NY 11794, USA\\
${}^*$ Center for Theoretical Physics of Complex Systems,\\
Institute for Basic Science (IBS), 55, Expo-ro, Yuseong-gu, Daejeon 34126, Republic of Korea\\

\vskip 0.5cm 

\email{olof.salberger@gmail.com, pramod23phys@gmail.com, korepin@gmail.com}

\vskip 0.5cm 

\end{center}


\abstract{
\noindent 
The Motzkin spin chain is a spin-1 model introduced in \cite{shor} as an example of a system exhibiting a high degree of quantum fluctuations whose ground state can be mapped to Motzkin paths that are generated with local equivalence moves. This model is difficult to solve in general but keeping just the height preserving local equivalence moves we show that the model becomes integrable which when projected to certain subspaces of the full Hilbert space is isomorphic to the spin-$\frac{1}{2}$ XXX chain. In fact in the full Hilbert space the system is akin to two non-interacting spin-$\frac{1}{2}$ XXX chains making the spectrum the same as the latter with the change coming in the degeneracy of the states. We then show that including the height-changing local-equivalence move is the same as introducing interactions in the above system.

}


\section{Introduction}
 
Entanglement is a feature that separates the quantum and classical worlds paving the way for quantum technology. One way of measuring this quantity is by computing entanglement entropy (EE). In this regard quantum many body systems described by local Hamiltonians gain importance. Restricting ourselves to $1D$ systems we know that gapped systems obey the area law \cite{gap3, gap4, gap1, gap2}. This is expected given the local nature of the interactions. However it is of interest to find local Hamiltonians that exhibit more quantum fluctuations with their EE scaling as either the logarithm or the volume of the system size \cite{ira, shor1}. More recently a spin-1 local, frustration-free Hamiltonian called the Motzkin spin chain \cite{shor, shor2} and a half-integer analog called the Fredkin spin chain \cite{fredkin} were shown to have unique ground states with the EE scaling as $\sqrt{n}$, where $n$ is the volume of the system. These models have since then been deformed to show phase transitions \cite{i1, i2}, generalized to possess new features using {\it symmetric inverse semigroups} \cite{p1, p2} and simplified to preserve translational invariance in \cite{caha}. 

The ground states of these models can be mapped to random walks in $2D$ such as the {\it Motzkin} and the {\it Dyck} walks and studied using techniques from enumerative combinatorics \cite{enum}. However the excited states in these systems are poorly understood. In this paper we take a step towards obtaining the spectrum of the Motzkin spin chain \cite{shor}. The bulk and boundary terms of the Hamiltonian are a sum of projectors and is given by
\begin{equation}
H_{Motzkin} = H_1 + \sum_{j=1}^{L-1}~\left[\hat{U}_j + \hat{D}_j + \epsilon \hat{F}_j\right] + H_L,
\end{equation}
where the operators $H_1$ and $H_L$ are the left and right boundary terms respectively and the bulk of the Hamiltonian is made of operators that project out the states shown in figures \ref{leM}, \ref{flatM}. Note that we have included a parameter $\epsilon>0$ in the bulk that projects out the last state shown in figure \ref{flatM}. In this paper we will analyze the case where $\epsilon = 0$ and observe that the resulting system is integrable and compute its spectrum for periodic boundary conditions using the coordinate Bethe ansatz \cite{cBethe, GaudinBook}. By turning on $\epsilon$ the system loses its symmetries and starts to `interact'. Thus we will call the case with $\epsilon=0$ as the {\it non-interacting Motzkin chain} or the {\it free Motzkin spin chain} and denote the corresponding Hamiltonian $H_{FM}$. 

Further details of the model are organized as follows. The setup of the model is described in section \ref{sec2} including its symmetries. The integrability of the model is shown in section \ref{sec22}. The spectrum of the model with periodic boundary conditions is discussed in section \ref{sec3}. Finally we show why turning on the $\epsilon$ parameter is equivalent to introducing interactions and study the algebra of operators of the Motzkin spin chain in section \ref{sec4}. We conclude with an outlook in section \ref{sec5}.

\section{The Hamiltonian $H_{FM}$}\label{sec2}

\begin{figure}[h!]
\captionsetup{width=0.8\textwidth}
\begin{center}
		\includegraphics[scale=0.8]{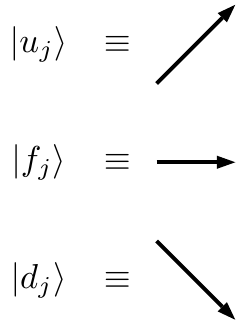} 
	\caption{\small Identifying the local Hilbert space with steps in the `x-y' plane. The links of the chain are indexed by $j$. }
\label{hilbM}
\end{center}
\end{figure}

Consider a spin 1 chain of length $L$ with the local Hilbert space located on the links. We denote the local basis states by $\{\ket{u_j}, \ket{f_j}, \ket{d_j}\}$ where $u$, $f$ and $d$ are used to abbreviate ``up'', ``flat'' and ``down'' respectively and $j$ an index for the links of the chain. This identification comes from the fact that states in this system can be mapped to paths in the `x-y' plane as done in \cite{shor} for example. So the state $\ket{u_j}$ maps to the $(1,1)$ direction, $\ket{f_j}$ maps to the $(1,0)$ direction and $\ket{d_j}$ maps to the $(1,-1)$ direction in the `x-y' plane as shown in figure \ref{hilbM}. 

\begin{figure}[h!]
\captionsetup{width=0.8\textwidth}
\begin{center}
		\includegraphics[scale=0.8]{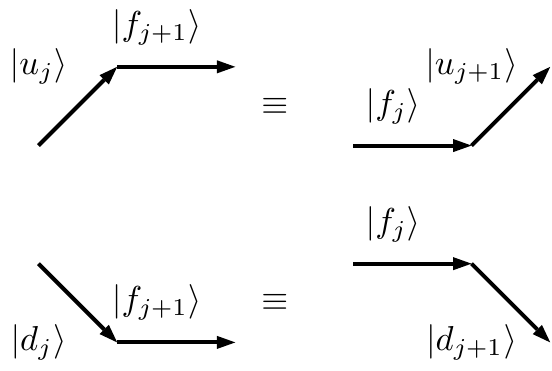} 
	\caption{\small The height preserving local equivalence moves of the non-interacting Motzkin spin chain. The local operators $\hat{U}_j$, $\hat{D}_j$ project out these states respectively. }
\label{leM}
\end{center}
\end{figure}

 The Hamiltonian for periodic boundary conditions is given by
\begin{equation}\label{hfm1}
H_{FM} = \sum_{j=1}^L~\left[\hat{U}_j + \hat{D}_j\right], 
\end{equation}
where the operators $\hat{U}_j$ and $\hat{D}_j$ are projectors to the states 
$$ \ket{u_j,~f_{j+1}} - \ket{f_j,~u_{j+1}}, $$
and 
$$ \ket{d_j,~f_{j+1}} - \ket{f_j,~d_{j+1}} $$
respectively. With the identification of the states with paths in the `x-y' plane we can interpret these states as local equivalence of paths as shown in figure \ref{leM}.
In the closed chain link $L+1$ is identified with 1.

At first glance the local equivalence moves of figure \ref{leM} indicate that the Hamiltonian is of the permutation type as the two moves interchange the $\ket{u_j}$ and $\ket{f_{j+1}}$ states and the $\ket{d_j}$ and $\ket{f_{j+1}}$ states respectively. This suggests that the Hamiltonian must be of the spin-$\frac{1}{2}$ XXX type but in two different subspaces of the full Hilbert space, namely the subspace where $\ket{d_j}=0,~\forall j\in\{1,\cdots, L\}$ and the subspace where $\ket{u_j}=0,~\forall j\in\{1,\cdots, L\}$. This is easily verified through equations as

\begin{eqnarray}
\hat{U}_j & = &  \ket{u_j,~f_{j+1}}\bra{u_j,~f_{j+1}} - \ket{u_j,~f_{j+1}}\bra{f_j,~u_{j+1}} \nonumber \\ 
& - & \ket{f_j,~u_{j+1}}\bra{u_j,~f_{j+1}} + \ket{f_j,~u_{j+1}}\bra{f_j,~u_{j+1}}, \nonumber \\ 
& = & \frac{1}{2}~\left[ 1^u_j1^u_{j+1} - u^x_ju^x_{j+1}- u^y_ju^y_{j+1}- u^z_ju^z_{j+1}\right],
\end{eqnarray}
with 
\begin{equation}
1^u = \left(\begin{array}{ccc} 1 & 0 & 0 \\ 0 & 1 & 0 \\ 0 & 0 & 0\end{array}\right),~ u^x = \left(\begin{array}{ccc} 0 & 1 & 0 \\ 1 & 0 & 0 \\ 0 & 0 & 0\end{array}\right), ~ u^y = \left(\begin{array}{ccc} 0 & -i & 0 \\ i & 0 & 0 \\ 0 & 0 & 0\end{array}\right), ~ u^z = \left(\begin{array}{ccc} 1 & 0 & 0 \\ 0 & -1 & 0 \\ 0 & 0 & 0\end{array}\right).
\end{equation}
Note that these are the spin-$\frac{1}{2}$ Pauli matrices embedded in three dimensional space. It just signifies the fact the subspace of the full Hilbert space with $\ket{d_j}=0,~\forall j\in\{1,\cdots, L\}$ is isomorphic to the spin-$\frac{1}{2}$ system with the local basis on each link $j$ spanned by $\ket{u_j}, \ket{f_j}$. We can also use the projector that commutes with $H_{FM}$ 
\begin{equation}
P_d = \prod_{j=1}^L~\left[\ket{f_j}\bra{f_j} + \ket{u_j}\bra{u_j} + \ket{u_j}\bra{d_j}\right],
\end{equation}
to go to the subspace where $\ket{d_j}=0,~\forall j\in\{1,\cdots, L\}$.  

In a similar manner we can obtain the operator for the other equivalence move as
\begin{equation}
\hat{D}_j = \frac{1}{2}~\left[ 1^d_j1^d_{j+1} - d^x_jd^x_{j+1}- d^y_jd^y_{j+1}- d^z_jd^z_{j+1}\right],
\end{equation}
with 
\begin{equation}
1^d = \left(\begin{array}{ccc} 0 & 0 & 0 \\ 0 & 1 & 0 \\ 0 & 0 & 1\end{array}\right),~ d^x = \left(\begin{array}{ccc} 0 & 0 & 0 \\ 0 & 0 & 1 \\ 0 & 1 & 0\end{array}\right), ~ d^y = \left(\begin{array}{ccc} 0 & 0 & 0 \\ 0 & 0 & -i \\ 0 & i & 0\end{array}\right), ~ d^z = \left(\begin{array}{ccc} 0 & 0 & 0 \\ 0 & 1 & 0 \\ 0 & 0 & -1\end{array}\right).
\end{equation}
This corresponds to the spin-$\frac{1}{2}$ subspace spanned by the local basis states  $\ket{d_j}, \ket{f_j}$. In this subspace $\ket{u_j}=0$ for each link $j$.

By defining the raising and lowering operators for the $\ket{u}$ and $\ket{d}$ states we can rewrite the Hamiltonian as 
\begin{eqnarray}
H_{FM} & = & \sum_{j=1}^L~\left[\frac{1^u_j1^u_{j+1}}{2} - u^+_ju^-_{j+1} - u^-_ju^+_{j+1} - \frac{u^z_j u^z_{j+1}}{2}\right] \nonumber \\ 
 & + & \sum_{j=1}^L~\left[\frac{1^d_j1^d_{j+1}}{2} - d^+_jd^-_{j+1} - d^-_jd^+_{j+1} - \frac{d^z_jd^z_{j+1}}{2}\right].
 \end{eqnarray}
 
 
 \paragraph{Symmetries of $H_{FM}$ -} 
From the form of the Hamiltonian it is clear that the total number of up steps, down steps and flat steps are conserved. That is the global operators $\sum\limits_{j=1}^L~\ket{u_j}\bra{u_j}$, $\sum\limits_{j=1}^L~\ket{d_j}\bra{d_j}$ and $\sum\limits_{j=1}^L~\ket{f_j}\bra{f_j}$ commute with $H_{FM}$. 

Apart from these number operators the projectors are invariant under the interchange of the up and down steps, that is we have the additional global symmetry, $\prod\limits_{j=1}^L~\left[\ket{u_j}\bra{d_j} + \ket{d_j}\bra{u_j}\right]$. Finally we also have translational invariance. 

Due to these symmetries the Hamiltonian block diagonalizes into invariant subspaces labelled by the number of up and down states, $u$ and $d$ respectively. Thus we can write any eigenstate in this system using the quantum numbers $(u, d)$.

 \section{Algebra of operators and Integrability} \label{sec22}

 This system can be mapped to the periodic Temperley-Lieb Hamiltonian \cite{TL} which is integrable. This is understood from the algebra of the two projectors, $\hat{U}_j$ and $\hat{D}_j$. Setting $\hat{U}_j + \hat{D}_j = \hat{e}_j$, we find
\begin{eqnarray}
\hat{e}_j\hat{e}_{j\pm 1}\hat{e}_j & = & \hat{e}_j, \label{e1}\\
\hat{e}_j^2 & = & 2\hat{e}_j, \label{e2}\\
\hat{e}_j\hat{e}_k & = & \hat{e}_k\hat{e}_j,~~|j-k|>1, \label{e3}
\end{eqnarray}
for all $j, k \in \{1,2,\cdots, L-1\}$. Additionally for the final generator $e_L$ we have due to periodicity
\begin{eqnarray}
\hat{e}_1\hat{e}_L\hat{e}_1 & = & \hat{e}_1, \label{e4}\\
\hat{e}_L\hat{e}_j\hat{e}_L & = & \hat{e}_L,~~j=1, L-1 \label{e5}\\
\hat{e}_L^2 & = & 2\hat{e}_L, \label{e6}\\
\hat{e}_j\hat{e}_L & = & \hat{e}_L\hat{e}_j,~~j\neq 1, L-1 \label{e7}
\end{eqnarray}
making the total algebra a {\it Periodic Temperley-Lieb algebra} (PTL) \cite{an10, an11, an30} which is one of the generalizations of the TL algebra \cite{TL, TL2}. 


Comparing these relations to the definition of the PTL algebra given in \cite{TL2}
\begin{equation}
e_i^2 = (q+q^{-1})e_i,
\end{equation} 
with the other relations being the same as in Eq. \ref{e1}- Eq. \ref{e7}, we find that $q=1$ in our case.  

Unlike the finite dimensional TL algebra the PTL algebra is infinite dimensional. However the spectrum of $H_{FM}$ will lie in a finite dimensional quotient. This is similar to the case of the XXZ and Potts models on the open and closed chains \cite{TL2}.

Furthermore we can also write down the $R$-matrix for $H_{FM}$ using the generators of the underlying PTL algebra \cite{btl}. We have
\begin{equation}
R_{j, j+1}(\lambda) = \hat{e}_j - \frac{\lambda+1}{\lambda},
\end{equation}
where $\lambda$ is the spectral parameter. Here $j$ denotes the index for the links of the chain and the operator $\hat{e}_j$ with support on the links $j$ and $j+1$. We can easily verify using the PTL algebra that this $R$-matrix satisfies the Yang-Baxter equation (YBE),
\begin{equation}
R_{j, j+1}(\lambda_1)R_{j+1, j+2}(\lambda_1+\lambda_2)R_{j, j+1}(\lambda_2) = R_{j+1, j+2}(\lambda_2)R_{j, j+1}(\lambda_1+\lambda_2)R_{j+1, j+2}(\lambda_1).  
\end{equation}
With this $R$-matrix we can use the method of algebraic Bethe ansatz to prove the integrability of $H_{FM}$ as provided in \cite{korbook} or in a more recent review \cite{abrev}.

These arguments show that $H_{FM}$ given in Eq. \ref{hfm1} is just the periodic TL Hamiltonian and is integrable. Hence we call it the {\it non-interacting Motzkin chain} or the {\it free Motzkin chain}.

\section{Spectrum of $H_{FM}$}\label{sec3}

As we have just noted the global symmetries of $H_{FM}$ imply that the invariant subspaces have the number of up and down steps fixed.  
So we set $u+d=r$, then the dimension of each invariant subspace is given by $\binomi{L}{r}2^r$ with $r= 0, \cdots, L$. The total dimension of the Hilbert space is then  
$$ \sum_{r = 0}^L~\binomi{L}{r}2^r = \sum_{r = 0}^L~\binomi{L}{r}2^r1^{L-r} = 3^L, $$
as expected. In order to find the eigenstates we make use of the fact that the states can be interpreted as paths in the `x-y' plane using the mapping shown in figure \ref{hilbM} and that they can be generated using the local equivalence moves shown in figure \ref{leM}. 

The eigenstates include degenerate ground states, that can be both product states and entangled states, and excited states. We will consider each of them separately.  

\paragraph{Product ground states -}  
 In this case $u+d=L$. As the Hamiltonian is a sum of projectors the ground state energy is 0. Thus the product ground states are the ones on which we cannot apply the local equivalence moves of the free Motzkin spin chain shown in figure \ref{leM}. For a chain of length $L$ the possibilities are listed in table \ref{tab1}. 

\begin{table}[h]
\centering
\begin{tabular}{|c|c|c|}
\hline
$u$ & $d$ & Number of inequivalent configurations\\
\hline
$L$  & 0 & $\binomi{L}{L}$\\
$L-1$ & 1 & $\binomi{L}{L-1}$ \\
$L-2$ & 2 & $\binomi{L}{L-2}$\\
\vdots & \vdots & \vdots\\
1 & $L-1$ & $\binomi{L}{1}$\\
0 & $L$ & $\binomi{L}{0}$\\
\hline
\end{tabular}
\caption{Possibilities for product ground states in a chain of length $L$ - Periodic boundary conditions.}
\label{tab1}
\end{table}

Each row of table \ref{tab1} represents one equivalence class of configuration under the local equivalence moves of the free Motzkin spin chain. Futhermore the binomial coefficient, $\binomi{L}{u} = \binomi{L}{d}$ gives the number of inequivalent configurations within each equivalence class. 
Thus for a given length $L$ we have 
$$ \sum_{u=0}^L~\binomi{L}{u} = 2^L,$$
product states with just $\ket{u}$ or $\ket{d}$ on each link. 

\begin{figure}[h!]
\captionsetup{width=0.8\textwidth}
\begin{center}
		\includegraphics[scale=0.7]{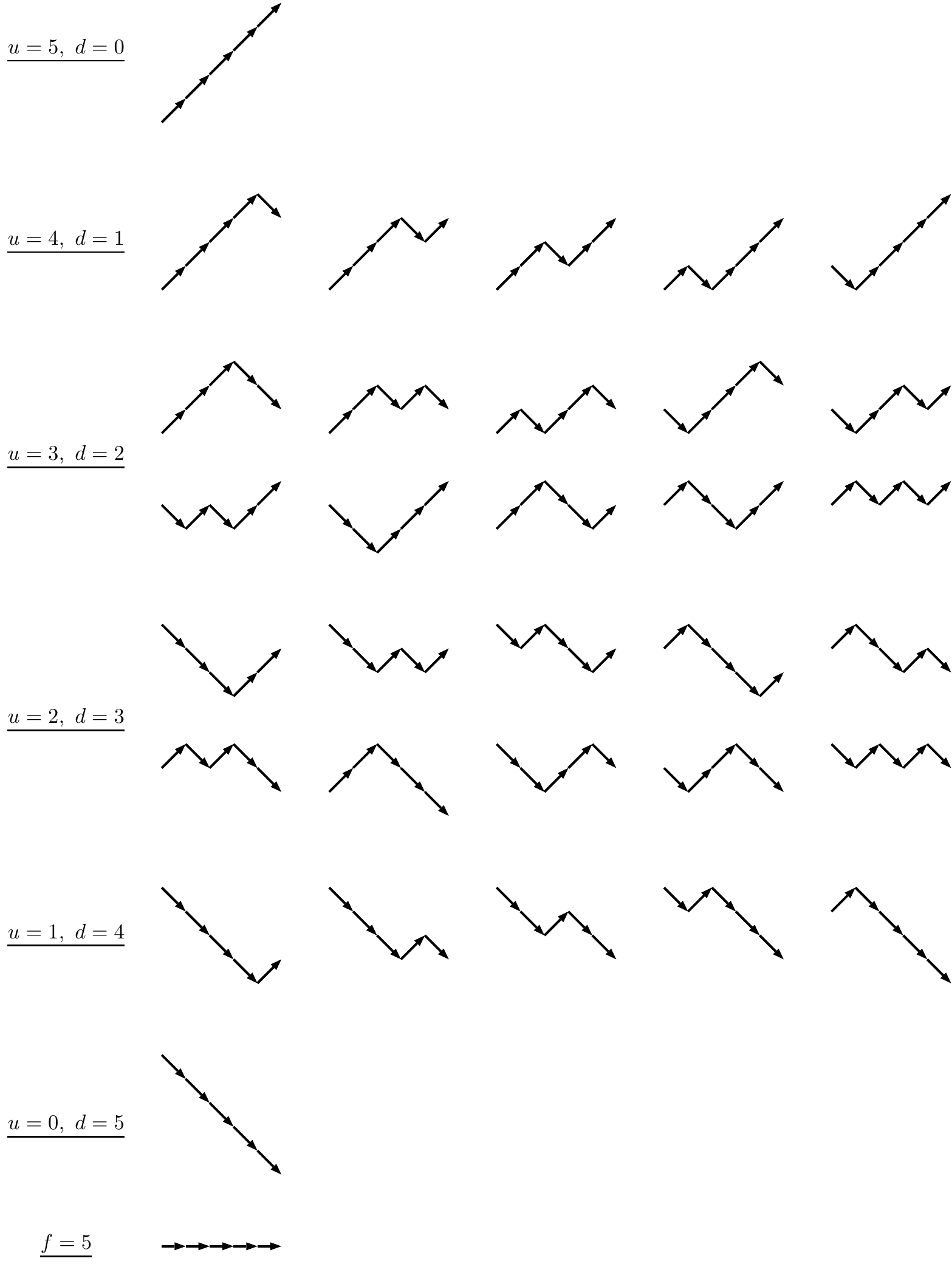} 
	\caption{\small The inequivalent product ground states in a chain of length $L=5$. We expect $2^5 +1 = 33$ states as shown in the figure.}
\label{pgc}
\end{center}
\end{figure}

Apart from these we also have the product state with flat steps on all the links or $u+d=0$. Thus the total number of ground states that are product states is given by
\begin{equation}
GSD_{product} =2^L+1.
\end{equation} 
For example in a chain of length, $L=5$ we have 33 states as shown in figure \ref{pgc}.

\paragraph{Entangled ground states -}  
When we include a flat step in a configuration otherwise made of the up and down steps, the local equivalence moves in figure \ref{leM} generate equivalent configurations. The equal weight superposition of these configurations form the entangled ground states. For a length $L$ chain, consider the $f=1$ sector. We then have $u+d=L-1$ and there are $2^{L-1}$ inequivalent entangled states as listed in table \ref{tab2}.  

\begin{table}[h]
\centering
\begin{tabular}{|c|c|c|c|}
\hline
$u$ & $d$ & $f$ & Number of inequivalent configurations \\
\hline
$L-2$ & 1 & 1 & $\binomi{L-1}{L-2}$ \\
$L-3$ & 2  & 1 & $\binomi{L-1}{L-3}$\\
\vdots & \vdots & \vdots & \vdots\\
1 & $L-2$ & 1 & $\binomi{L-1}{1}$\\
0 & $L-1$ & 1 & $\binomi{L-1}{0}$\\
\hline
\end{tabular}
\caption{Entangled ground states in a closed chain of length $L$ with one flat step.}
\label{tab2}
\end{table}

\begin{figure}[h!]
\captionsetup{width=0.8\textwidth}
\begin{center}
		\includegraphics[scale=0.7]{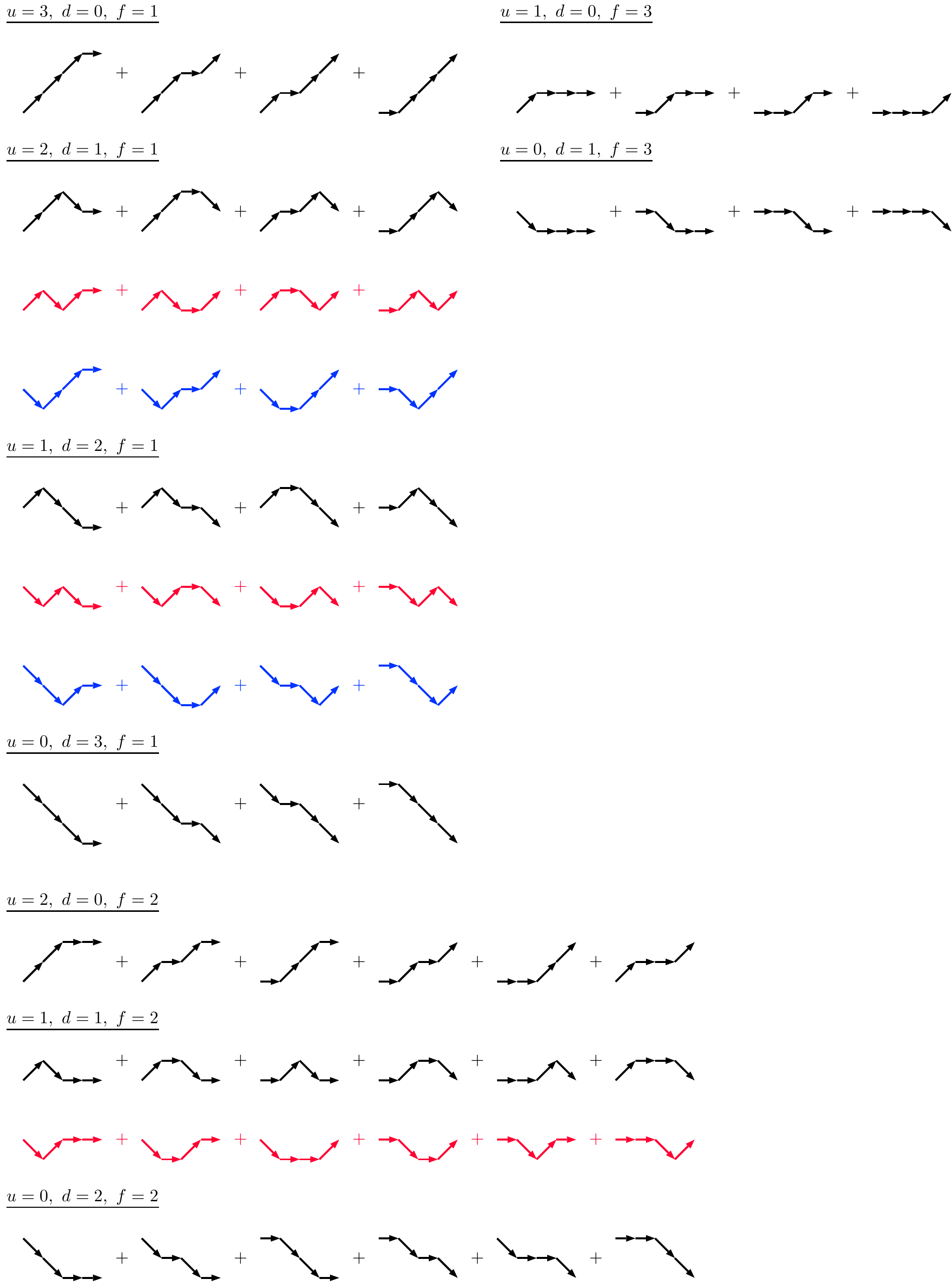} 
	\caption{\small The inequivalent entangled ground states in a chain of length $L=4$. We expect $2^4-2 = 14$ states as shown in the figure.}
\label{entg}
\end{center}
\end{figure}

In general for a sector with $f$ flat steps we have $2^{L-f}$ inequivalent entangled ground states giving us 
\begin{equation}
GSD_{entangled} = \sum_{f=1}^{L-1}~2^{L-f} = 2^L-2.
\end{equation}
 The normalization of each of these entangled ground states is $\sqrt{\binomi{L}{f}}$. We illustrate the entangled ground states for $L=4$ in figure \ref{entg}.

Thus the ground state degeneracy (GSD) of the free Motzkin chain is 
\begin{equation}
GSD = GSD_{product} + GSD_{entangled} = 2^{L+1}-1.
\end{equation} 

\subsection*{Excited States} 

To obtain the excited states we start with the product ground state where all the steps are flat,
$$ \ket{(0,0)} = \ket{f_1, f_2,\cdots, f_L},$$
 and flip each step into either an up step or a down step. We will write down the one-particle and the two-particle states using $\ket{(0,0)}$ as a reference state and employ the coordinate Bethe ansatz technique \cite{cBethe, GaudinBook}. The general $r$-particle state can be obtained easily once we know the expressions for the two-particle states as this system is integrable. 

\paragraph{Single particle excitations - } 
For the one particle states we have two possibilities, either $u=1, d=0$ or $u=0, d=1$. The states can be obtained using the coordinate Bethe ansatz \cite{cBethe} and have energy 
\begin{equation}
E(1,0; k_1) = E(0,1; k_1) = 2\left[1-\cos{k_1}\right]
\end{equation}
where $k_1$ is the momentum of the particle. As the system is invariant under the global interchange of up and down steps the energies for the two states are the same and we only need to obtain one of the eigenvectors.
The state is just the plane wave state given by 
\begin{equation}
\ket{(1,0; k_1)} = \frac{1}{\sqrt{L}}\times\sum_{n_1=1}^{L}~ e^{ik_1n_1}\ket{n_1},
\end{equation}
where $n_1$ denotes the position of the up step in the length $L$ chain filled with $L-1$ flat steps in the remaining links. By imposing periodic boundary conditions on the state the momenta are given by
\begin{equation}
k_1 = \frac{2\pi m}{L},~~m\in\{0,1,\cdots, L-1\}.
\end{equation}
Thus we have $L$ different one particle states as expected.
Note that for $k_1=0$ we obtain $E(1,0; 0)=E(0,1; 0)=0$ which are ground states for $H_{FM}$. The states are then equal weight superpositions of $\ket{n_1}$. 

A similar expression for the state $\ket{(0,1; k_1)}$ is obtained by applying the global symmetry interchanging the up and down steps. Thus we have the same one particle states known as the 1 magnon states of the spin-$\frac{1}{2}$ XXX chain with only an increase in the degeneracy by a factor of 2. 

\paragraph{Two-particle excitations -}  
In this case $u+d=2$ giving us three possibilities as listed in table \ref{tab3}. The states $\ket{(2,0; k_1, k_2)}$ and $\ket{(0,2; k_1, k_2)}$ are just the two-particle states of the spin-$\frac{1}{2}$ XXX chain. Their energy is given by
\begin{equation}
E(2,0; k_1, k_2)=E(0,2; k_1, k_2) = 2\left[2 - \cos{k_1} - \cos{k_2}\right],
\end{equation}
with $k_1$ and $k_2$ denoting the momenta of the two particles. The state is given by
\begin{equation}\label{2-part}
\ket{(2,0; k_1, k_2)} = \frac{\sqrt{2}}{\sqrt{L(L-1)}}\times \sum_{n_2>n_1}~f(n_1, n_2) \ket{n_1, n_2},
\end{equation}
where 
\begin{equation}
f(n_1, n_2) = e^{i\left(k_1n_1 + k_2n_2 + \frac{\theta_{12}}{2}\right)} + e^{i\left(k_2n_1 + k_1n_2 + \frac{\theta_{21}}{2}\right)},
\end{equation}
and $\theta_{12} = \theta = -\theta_{21}$, 
\begin{equation}\label{theta}
e^{i\theta} = - e^{i\left(k_1-k_2\right)}\left[\frac{e^{-ik_1} + e^{ik_2} - 2}{e^{ik_1} + e^{-ik_2} - 2}\right].
\end{equation}
The numbers $n_1$ and $n_2$ denote the positions of the two up steps and the remaining links are filled with flat steps. The periodic boundary conditions constrain the momenta to satisfy
\begin{equation} \label{mo2}
k_1+k_2 = \frac{2\pi}{L}\left(m_1+m_2\right),~~0\leq m_1\leq m_2\leq L-1.
\end{equation}
These are the {\it Bethe equations} and they have three kinds of solutions. When $k_1=0$ we have $L$ solutions for $k_2$ which are the one-magnon states discussed earlier. We have $\frac{(L-3)(L-2)}{2}$ two particle states that are the superposition of two one-magnon states with momenta satisfying $|m_2 - m_1| \geq 2$.  Finally we have $L-3$ bound state solutions with momenta satisfying $(m_1, m_2) \in \{0, \pm 1\} ~\textrm{mod}~L$.
Thus the total is seen to be $\frac{L(L-1)}{2}$, the dimension of the two-particle sector. 

Once again note that for $k_1=k_2=0$ we have $E(2,0; 0,0) = E(0,2;0, 0) = 0$ and the state $\ket{(2,0;0,0)}$ is just the ground state as an equal weight superposition of $\ket{n_1, n_2}$. 

The other possibility of $\ket{(0, 2; k_1, k_2)}$ can be obtained from $\ket{(2, 0; k_1, k_2)}$ by replacing the up steps with down steps exploiting the global symmetry interchanging the up and down steps. 

\begin{table}[h]
\centering
\begin{tabular}{|c|c|c|}
\hline
$u$ & $d$ & Dimension of the sector \\
\hline
2 & 0 & $\binomi{L}{2}\times\binomi{2}{2}$ \\
1 & 1  & $\binomi{L}{2}\times\binomi{2}{1}$\\
0 & 2 & $\binomi{L}{2}\times\binomi{2}{0}$\\
\hline
\end{tabular}
\caption{Possibilities for two-particle excitations in a chain of length $L$ with $L-2$ flat steps.}
\label{tab3}
\end{table}

Next we look at the third possible configuration where $u=1, d=1$ giving the state $\ket{(1,1; k_1, k_2)}$. We have two inequivalent configurations in this class given by either the up step followed by the down step or the down step followed by the up step. These states continue to have the same energy as the other two-particle states given by
\begin{equation}\label{e11}
E(1, 1; k_1, k_2) = 2\left[2 - \cos{k_1} - \cos{k_2}\right].
\end{equation}
We can show that this is indeed the case by using the coordinate Bethe ansatz for the state $\ket{(1,1; k_1, k_2)}$. The state $\ket{(1,1; k_1, k_2)}$ takes the same form given by Eq. \ref{2-part} with $n_1$ and $n_2$ now denoting the positions of the up and down steps or down and up steps respectively. 
To obtain the action of $H_{FM}$ on $\ket{(1,1; k_1, k_2)}$ we need
\begin{equation}
\sum_{j=1}^L~\frac{1^u_j1^u_{j+1}}{2} \ket{n_1, n_2} = \left(\frac{L-2}{2}\right)~\ket{n_1, n_2},
\end{equation}
\begin{equation}
\sum_{j=1}^L~\frac{1^d_j1^d_{j+1}}{2} \ket{n_1, n_2} = \left(\frac{L-2}{2}\right)~\ket{n_1, n_2},
\end{equation}
\begin{equation}
\sum_{j=1}^L~ u^+_ju^-_{j+1} \ket{n_1, n_2} = \left\{\begin{array}{l} \ket{n_1 - 1, n_1 +1},~~\textrm{when}~n_2=n_1+1 \\ \ket{n_1-1, n_2},~~ n_2>n_1+1\end{array} \right.
\end{equation}
\begin{equation}
\sum_{j=1}^L~ d^+_jd^-_{j+1} \ket{n_1, n_2} = \left\{\begin{array}{l} 0,~~\textrm{when}~n_2=n_1+1 \\ \ket{n_1, n_2-1},~~ n_2>n_1+1\end{array} \right.
\end{equation}
\begin{equation}
\sum_{j=1}^L~ u^-_ju^+_{j+1} \ket{n_1, n_2} = \left\{\begin{array}{l} 0,~~\textrm{when}~n_2=n_1+1 \\ \ket{n_1+1, n_2},~~ n_2>n_1+1\end{array} \right.
\end{equation}
\begin{equation}
\sum_{j=1}^L~ d^-_jd^+_{j+1} \ket{n_1, n_2} = \left\{\begin{array}{l} \ket{n_1, n_1 +2},~~\textrm{when}~n_2=n_1+1 \\ \ket{n_1, n_2+1},~~ n_2>n_1+1\end{array} \right.
\end{equation}
\begin{equation}
\sum_{j=1}^L~ \frac{u^z_j}{2}\frac{u^z_{j+1}}{2} \ket{n_1, n_2} = \left\{\begin{array}{l} \left(\frac{L-3}{4} - \frac{1}{4}\right)\ket{n_1, n_1 +1},~~\textrm{when}~n_2=n_1+1 \\ \left(\frac{L-4}{4} - \frac{1}{2}\right)\ket{n_1, n_2},~~ n_2>n_1+1\end{array} \right.
\end{equation}
\begin{equation}
\sum_{j=1}^L~ \frac{d^z_j}{2}\frac{d^z_{j+1}}{2} \ket{n_1, n_2} = \left\{\begin{array}{l} \left(\frac{L-3}{4} - \frac{1}{4}\right)\ket{n_1, n_1 +1},~~\textrm{when}~n_2=n_1+1 \\ \left(\frac{L-4}{4} - \frac{1}{2}\right)\ket{n_1, n_2},~~ n_2>n_1+1\end{array} \right. .
\end{equation}

On applying $H_{FM}$ on $\ket{(1,1; k_1, k_2)}$ we have the following form
\begin{eqnarray}
H_{FM} \ket{(1,1; k_1, k_2)} & = & E(1,1; k_1, k_2)  \left[\sum_{n_2 > n_1+1}f(n_1, n_2) \ket{(n_1, n_2)} \right. \nonumber \\
   & + & \left. \sum_{n=1}^Lf(n, n+1) \ket{(n, n+1)}\right].
\end{eqnarray}
Substituting the coordinate Bethe ansatz for the coefficients,
\begin{equation}
f(n_1, n_2) = e^{i\left(k_1n_1 + k_2n_2 + \frac{\theta_{12}}{2}\right)} + e^{i\left(k_2n_1 + k_1n_2 + \frac{\theta_{21}}{2}\right)},
\end{equation}
and comparing terms we obtain the energy as given in Eq. \ref{e11}. The scattering angle is given by Eq. \ref{theta} and the momenta by Eq. \ref{mo2}. And once again for $k_1=k_2=0$ we have $E(1,1;k_1, k_2)=0$ and there are two entangled ground states.

Thus the two-particle states of $H_{FM}$ are the same as the two-particle states of the spin-$\frac{1}{2}$ XXX chain but with a degeneracy of $ 2L(L-1)-4$. 

\paragraph{A general $r$-particle excited state -} 
In these states $u+d =r$ and there are $L-r$ flat steps. The possibilities for the different inequivalent configurations are listed in table \ref{tab4} giving a total of $\binomi{L}{r}\times 2^r$ states as noted earlier. 

\begin{table}[h]
\centering
\begin{tabular}{|c|c|c|}
\hline
$u$ & $d$ & Number of configurations \\
\hline
$r$ & 0 & $\binomi{L}{r}\times\binomi{r}{r}$ \\
$r-1$ & 1  & $\binomi{L}{r}\times\binomi{r}{r-1}$\\
$r-2$ & 2 & $\binomi{L}{r}\times\binomi{r}{r-2}$\\
\vdots & \vdots & \vdots \\
1 & $r-1$ & $\binomi{L}{r}\times\binomi{r}{1}$\\
0 & $r$ &  $\binomi{L}{r}\times\binomi{r}{0}$\\
\hline
\end{tabular}
\caption{Possibilities for $r$-particle excitations in a chain of length $L$ with $L-r$ flat steps.}
\label{tab4}
\end{table}
These states have energy 
\begin{equation}
E(u,d;k_1,\cdots,k_r) = 2\left[r - \sum_{j=1}^r~\cos{k_j}\right],
\end{equation}
with $k_j$ the momenta of the $r$ particles.

The eigenstates take the same form as the $r$-particle excitations in the spin-$\frac{1}{2}$ XXX chain,
\begin{equation}
\ket{(u, r-u;k_1,\cdots,k_r)} = \sum_{L\geq n_r>n_{r-1}>\cdots>n_2>n_1\geq1}~ f(n_1, n_2, \cdots, n_r)\ket{(n_1, n_2, \cdots, n_r)},
\end{equation}
with
\begin{equation}
f(n_1, n_2, \cdots, n_r) = \sum_{P\in S_r}~e^{\left[i\sum_{j=1}^r~k_{P(j)}n_j + \frac{i}{2}\sum_{l<j}~\theta_{P(l)P(j)}\right]},
\end{equation}
where $S_r$ denotes the permutation group of $r$ elements and $P(j)$ is an element of $S_r$ that sends index $j$ to some number between 1 and $r$. And the scattering angles are given by 
\begin{equation}
e^{i\theta_{jl}} = -e^{i(k_j-k_l)}\left[\frac{e^{-ik_j} + e^{ik_l} - 2}{e^{ik_j} + e^{-ik_l} - 2}\right].
\end{equation}
The momenta of the $r$ particles is again determined by the periodicity of the wavefunction and is given by
\begin{equation}
\sum_{j=1}^r~ k_j = \frac{2\pi}{L}\left(\sum_{j=1}^r~m_j\right),~~ 0\leq m_1\leq m_2\leq\cdots m_r\leq L-1.
\end{equation} 
Thus the $r$-particle states are identical to the spin-$\frac{1}{2}$ XXX chain with the difference being that these states obtain a degeneracy of $2^r\times\left[\binomi{L}{r}-1\right]$. The remaning $2^r$ states are the entangled ground states which occur when all the momenta $k_j=0$ as we saw in the one-particle and the two-particle cases. 

This completes the analysis of the full spectrum of our model which can be thought of as two decoupled spin-$\frac{1}{2}$ XXX chains and hence this is like the free part of the full Motzkin spin chain \cite{shor}. 

\section{The flat moves as interactions}\label{sec4}

\begin{figure}[h!]
\captionsetup{width=0.8\textwidth}
\begin{center}
		\includegraphics[scale=0.8]{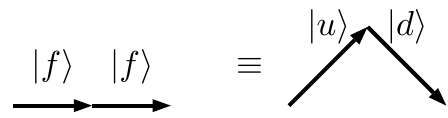} 
	\caption{\small The interaction inducing flat move. }
\label{flatM}
\end{center}
\end{figure}

The Motzkin spin chain \cite{shor} contains one more local equivalence move which we call the flat move as shown in figure \ref{flatM}.

This move changes the height of the paths as is evident from the figure \ref{flatM} and is like an interaction term which can be seen explicitly from its expression
\begin{eqnarray}
\hat{F}_j & = & \left[\ket{f_j, f_{j+1}} - \ket{u_j, d_{j+1}}\right]\left[\bra{f_j, f_{j+1}} - \bra{u_j, d_{j+1}}\right], \nonumber \\
 & = & \frac{1}{4}\left[1^u_j1^u_{j+1} + 1^u_j1^d_{j+1} - u^z_j1^d_{j+1} - 1^u_jd^z_{j+1} - u^z_jd^z_{j+1} - u^z_ju^z_{j+1} - u^z_j - u^z_{j+1}\right] \nonumber \\ & - & \frac{1}{2}\left[u^x_jd^x_{j+1} - u^y_jd^y_{j+1}\right].
 \end{eqnarray} 

The inclusion of the flat moves as an interaction to the free Motzkin chain drastically modifies the PTL algebra structure of the up and down moves discussed earlier. We will now write down the relations for this modified algebra and identify subalgebras that are isomorphic to known structures.
 
 As before we set $\hat{U}_j + \hat{D}_j = \hat{e}_j$ and also set $\hat{F}_j = \hat{f}_j$. To give a simplified form for the relations we require some additional definitions of operators,
 \begin{eqnarray} 
 \hat{f}_{j}\hat{f}_{j+1}\hat{f}_{j} & = & \hat{g_1}_{j}, \label{g1}\\
 \hat{e}_{j}\hat{f}_{j+1}\hat{e}_{j} & = & \hat{g_2}_{j}, \label{g2}\\
  \hat{e}_{j}\hat{f}_{j+1}\hat{f}_{j}\hat{f}_{j+1}\hat{e}_{j} & = & \hat{g_3}_{j} \label{g3}, 
  \end{eqnarray}
  which will be shown to be projectors acting as partial identities, along with the following nilpotent operators,
\begin{eqnarray}
  \hat{e}_{j}\hat{f}_{j+1}\hat{f}_{j} & = & \hat{h_1}_{j}, \label{h1}\\
   \hat{e}_{j}\hat{e}_{j+1}\hat{f}_{j} & = & \hat{h_2}_{j}, \label{h2}\\
 \hat{f}_{j}\hat{e}_{j+1}\hat{e}_{j} & = & \hat{h_3}_{j}, \label{h3}\\
 \hat{f}_{j}\hat{f}_{j+1}\hat{e}_{j} & = & \hat{h_4}_{j}. \label{h4}
 \end{eqnarray}
The relations can now be summarized as
\begin{eqnarray}
\hat{e}_j\hat{e}_{j+1}\hat{e}_{j} & = & \hat{e}_j, \\
\hat{f}_{j}\hat{e}_{j+1}\hat{f}_{j} & = & \hat{f}_j, \\
\hat{h_2}_j\hat{h_3}_j & = & \hat{g_2}_j, \\ 
\hat{h_3}_j\hat{h_2}_j & = & \hat{f}_j, \\ 
\hat{h_1}_j\hat{h_3}_j = \hat{h_1}_j\hat{h_4}_j & = & \hat{h_2}_j\hat{h_4}_j = \hat{g_3}_j, \\
\hat{h_4}_j\hat{h_1}_j = \hat{h_4}_j\hat{h_2}_j & = & \hat{h_3}_j\hat{h_1}_j = \hat{g_1}_j, 
\end{eqnarray}

We also have
\begin{equation}
\begin{array}{ccc}
\hat{h_1}_j\hat{g_1}_j  =  \hat{h_1}_j, ~~ &  \hat{g_2}_j\hat{h_1}_j  =  \hat{h_1}_j, ~~  & \hat{g_3}_j\hat{h_1}_j  =  \hat{h_1}_j,  \\
\hat{h_2}_j\hat{g_1}_j  =  \hat{h_2}_j, ~~ &  \hat{g_2}_j\hat{h_2}_j  =  \hat{h_2}_j, ~~  & \hat{g_3}_j\hat{h_2}_j  =  \hat{h_1}_j,   \\
\hat{g_1}_j\hat{h_3}_j  =  \hat{h_3}_j, ~~ &  \hat{h_3}_j\hat{g_2}_j  =  \hat{h_3}_j, ~~  & \hat{h_3}_j\hat{g_3}_j  =  \hat{h_4}_j,   \\
\hat{g_1}_j\hat{h_4}_j  =  \hat{h_4}_j, ~~ &  \hat{h_4}_j\hat{g_2}_j  =  \hat{h_4}_j, ~~  & \hat{h_4}_j\hat{g_3}_j  =  \hat{h_4}_j.  
\end{array}
\end{equation}

The subsets $\{\hat{g_2}_j, \hat{f}_j, \hat{h_2}_j, \hat{h_3}_j\} $ and $\{\hat{g_1}_j, \hat{g_3}_j, \hat{h_1}_j, \hat{h_4}_j\} $ form subalgebras that are isomorphic to the symmetric inverse semigroup $\mathcal{S}^2_1$ which is generated by $x_{i,j}~;~i,j\in\{1,2\}$ with the composition rule $x_{i,j}x_{k,l} = \delta_{jk}x_{i,l}$.  

The other relations are obtained by interchanging $j$ and $j+1$ in Eqs. \ref{g1} - \ref{h4}.

\section{Outlook}\label{sec5}

We can change the boundary conditions for this system and carry out the same analysis. In particular we can include the boundary terms of the original Motzkin spin chain given by
\begin{equation}
H_1 = \ket{d_1}\bra{d_1},~~H_L = H_1 = \ket{u_L}\bra{u_L}.
\end{equation}
The system then loses the global symmetry that interchanges the up and down states. We then need to check if the system retains its integrability and use the Bethe ansatz technique for open chains \cite{btl}, \cite{sklb}. Apart from this we can also find integrable boundaries for this system \cite{ppkor}.

We have only worked with the {\it colorless} Motzkin spin chain in this paper and it is straightforward to generalize this analysis to the colored case \cite{shor}. 

We also plan to further study the spectrum of the full Motzkin spin chain by taking a more detailed look at its operator algebra. 

\section*{Acknowledgements}
PP was supported by the Institute of Basic Science in Korea (IBS-R024-Y1, IBS-R024-D1).

\end{document}